\documentclass[pra,twocolumn, showpacs, superscriptaddress,preprintnumbers,amsmath,amssymb, longbibliography]{revtex4}  % for review and submission

\usepackage{hyperref}

\usepackage{graphicx}  % needed for figures
\usepackage{dcolumn}   % needed for some tables
\usepackage{bm}        % for math
\usepackage{amssymb}   % for math
\usepackage{natbib}

\usepackage{color}

\usepackage{bm}

\begin{document}

\title{Arrested relaxation in an isolated molecular ultracold plasma}

\author{R. Haenel}
\affiliation{Department of Physics \& Astronomy, University of British Columbia, Vancouver, BC V6T 1Z3, Canada}
\author{ M. Schulz-Weiling}
\affiliation{Department of Physics \& Astronomy, University of British Columbia, Vancouver, BC V6T 1Z3, Canada}
\author{J. Sous}  
\affiliation{Department of Physics \& Astronomy, University of British Columbia, Vancouver, BC V6T 1Z3, Canada}
\affiliation{Stewart Blusson Quantum Matter Institute, University of British Columbia, Vancouver, British Columbia, V6T 1Z4, Canada}
\author{H. Sadeghi} 
\affiliation{Department of Chemistry, University of British Columbia, Vancouver, BC V6T 1Z3, Canada}
\author{M. Aghigh} 
\affiliation{Department of Chemistry, University of British Columbia, Vancouver, BC V6T 1Z3, Canada}
\author{L. Melo}
\affiliation{Department of Chemistry, University of British Columbia, Vancouver, BC V6T 1Z3, Canada}
\author{J. S. Keller}
 \altaffiliation[Permanent address: ] {Department of Chemistry, Kenyon College, Gambier, Ohio 43022 USA}
 \affiliation{Department of Chemistry, University of British Columbia, Vancouver, BC V6T 1Z3, Canada}
\author{E. R. Grant}
\email[Author to whom correspondence should be addressed. Electronic mail:  ]
{edgrant@chem.ubc.ca}
\affiliation{Department of Physics \& Astronomy, University of British Columbia, Vancouver, BC V6T 1Z3, Canada}
\affiliation{Department of Chemistry, University of British Columbia, Vancouver, BC V6T 1Z3, Canada}

\begin{abstract}

Spontaneous avalanche to plasma splits the core of an ellipsoidal Rydberg gas of nitric oxide.  Ambipolar expansion first quenches the electron temperature of this core plasma.  Then, long-range, resonant charge transfer from ballistic ions to frozen Rydberg molecules in the wings of the ellipsoid  quenches the centre-of-mass ion/Rydberg molecule velocity distribution.  This sequence of steps gives rise to a remarkable mechanics of self-assembly, in which the kinetic energy of initially formed hot electrons and ions drives an observed separation of plasma volumes.  These dynamics adiabatically sequester energy in a reservoir of mass transport, starting a process that anneals separating volumes to form an apparent glass of strongly coupled ions and electrons.  Short-time electron spectroscopy provides experimental evidence for complete ionization.  The long lifetime of this system, particularly its stability with respect to recombination and neutral dissociation, suggests that this transformation affords a robust state of arrested relaxation, far from thermal equilibrium.

\end{abstract}

\pacs{05.65.+b, 52.27.Gr, 32.80.Ee, 37.10.Mn, 34.80.Pa}

\maketitle

\section{Introduction}

Atomic and molecular gases driven to produce ultracold plasmas yield isolated systems with properties that can provide an important gauge of collision and transport under strongly coupled conditions \cite{Deutsch}.  Strong coupling occurs in a plasma when the average inter-particle potential energy, exceeds the average kinetic energy \cite{Ichimaru}.  This condition causes the formation of extremely non-ideal charged-particle systems that have fundamentally altered, fluid-like properties, which give rise to states of structural and dynamical order.  The physics of strong coupling play important but incompletely understood roles governing the dynamics of natural plasmas over a wide range of length scales \cite{Killian_Phys_rept}. 

Ultracold plasmas provide an ideal regime in which to study strong coupling in the laboratory.   They form optically thin, mm-sized objects that can be imaged in three-dimensions \cite{Killian_wave,Mcquillen,Cummings,MSW_bifur}. They evolve in a dynamics of particle-particle interactions that are easily probed by their response to external forces.  Yet such systems self-organize to form spatial and temporal structures \cite{Morrison_shock}.   Coulomb equations of motion scale \cite{Morrison2012}, so observations on cold rarified systems extend to represent the classical fluid properties of soft matter over a very wide range of temperature and density.  

Formed in states far from equilibrium, the ultracold plasmas created in a magneto-optical trap (MOT) generally relax in a manner well described by collisional rate processes and hydrodynamics in a quasi-thermal regime of classical mechanics \cite{Pohl2003,Killian_Phys_rept,Laha,PPR,PVS}, in which disorder-induced heating and three-body recombination lead to electron heating and plasma dissipation.  

Ultracold plasmas prepared in a supersonic molecular beam behave differently.  At beam densities, we have found that cold Rydberg gases of nitric oxide evolve to form a long-lived ultracold plasmas with hydrodynamic properties that appear to reflect anomalously low electron temperatures \cite{Morrison2009,Saquet2011,Sadeghi:2012}.  

The present work describes particular conditions under which this higher-density evolution partitions energy to form a spatially correlated, strongly coupled plasma.   Here, the  laser-crossed molecular beam geometry creates a density gradient in which spontaneous avalanche to plasma splits the core of an ellipsoidal Rydberg gas of nitric oxide.  Ambipolar expansion, accompanied by quenches, first of the electron temperature and then of the centre-of-mass ion velocity distribution, gives rise to a remarkable process of self-assembly in which the initial energy hot electrons drives the separation of plasma volumes.  The long lifetime of this system, particularly its stability with respect to recombination and neutral dissociation, suggests that bifurcation anneals these volumes to yield strongly coupled electrons and ions in a state of arrested relaxation, far from conventional thermal equilibrium.

\nopagebreak[4]

\section{A molecular ultracold plasma}

\subsection{Initial phase-space distribution}

\begin{figure}
    \begin{center}
        \includegraphics[width=.5\textwidth]{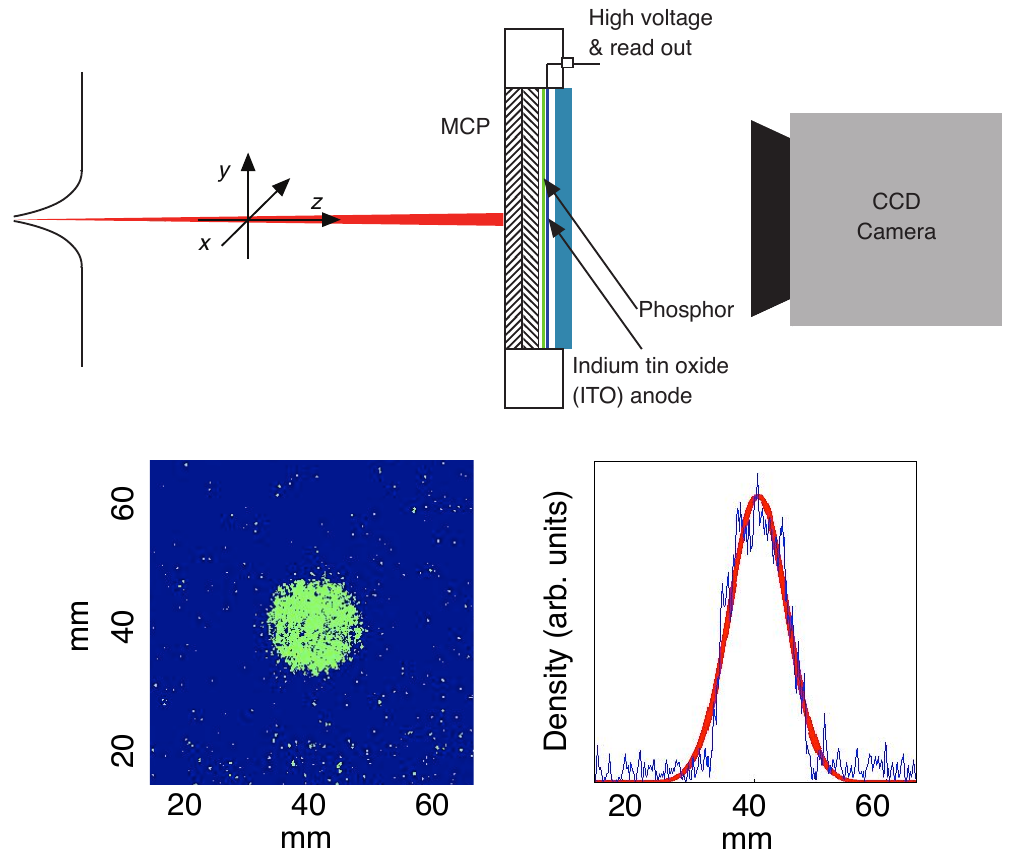}
    \end{center}
    \caption{Top: Experimental setup for the imaging of the molecular beam. The beam enters the experiment chamber through the skimmer and impacts on a MCP stack after a free-flight of 468mm. Free electrons are created, multiplied and viewed as photons by a CCD camera. Bottom-Left: CCD response of molecular beam impacting on phosphor screen detector (at a distance of 468 mm from the skimmer) without laser light present; Bottom-Right: Vertical summation over image pixel and fit, fitted Gaussian width is 4.35mm.} 
    \label{fig:imaging}
\end{figure}

Expanding from a high-pressure pulsed nozzle, the gas in a supersonic molecular beam attains a sub-Kelvin temperature in a local frame that moves with high velocity in the laboratory.  This differs substantially from the conditions encountered after the laser cooling of a gas of atoms in the magneto-optical trap (MOT) environment normally used in studies of ultracold plasmas \cite{Killian_Phys_rept}.  The supersonic beam environment offers two distinct advantages for the study of ultracold plasma physics:  Charge-particle densities in beams can exceed those attainable in a MOT by orders of magnitude \cite{tutorial}.  This medium also supports the formation of a \emph{molecular} ultracold plasma, which introduces interesting and potentially important new degrees of freedom \cite{Morrison2008,Sadeghi:2014}.  

The present experiment focuses on the plasma that evolves from a state-selected Rydberg gas, prepared by the double-resonant excitation of nitric oxide entrained 1:10 in a skimmed supersonic beam of helium from a stagnation pressure of 5 bar.  This jet travels 35 mm to transit a 1 mm diameter skimmer separating the source and experimental chambers of a differentially pumped vacuum system.  

The gas enters the skimmer with a translational temperature of 0.6 K and a mean free path on the order of 1 mm, enabling the use of a point source approximation to accurately describe the $z$-propagation of the gas in the experimental chamber as a molecular beam.  Figure \ref{fig:imaging} shows the $x,y$ density distribution of this supersonic molecular beam observed after propagating 468 mm from the skimmer to strike a dual multichannel plate imaging detector with a phosphor-screen anode \cite{Stephen}. 

In a typical experiment, we intercept this beam 150 mm from the skimmer where it presents a local peak ground state nitric oxide density of $1.6 \times 10^{14}$ cm$^{-3}$.  The translational temperature in the longitudinal direction remains, $T_z= 0.6$ K.  But, phase-space cooling associated with divergence,  $\gamma_{x}(0)$ and $\gamma_{y}(0)$, in the transverse degrees of freedom, cools motion in the $x$ and $y$ dimensions to less than 5 mK \cite{tutorial}.  The Gaussian width of the beam increases by this point to 3 mm fwhm ($\sigma_x=1.27$ mm). 

 Crossing this molecular beam in the $x$ direction with a 0.5 mm diameter pulsed laser beam ($\omega_1$, Gaussian width, $\sigma_y = \sigma_z = 0.2$ mm) marks a 6:1 ellipsoidal volume of nitric oxide with the hydrodynamic properties given in Table \ref{tab:beam}.   \setlength{\tabcolsep}{.8 em}
\begin{table}[h!]
 \caption{Thermodynamic and hydrodynamic properties of 6:1 and 2:1 ellipsoidal volumes defined by excitation geometries in which the $\omega_1 + \omega_2$ laser beams cross the molecular beam at distances 150 mm and 75 mm beyond the skimmer.  The NO density on axis remains constant over this interval.} 
 \label{tab:beam}
 \vspace{5 pt}
 \begin{tabular}{clll} 
 &150 mm  & & 75 mm \\
 \toprule 
  $\sigma_x(0)$ & 1.27 mm & & 0.75 mm\\
  $\sigma_y(0)$ & 0.2 mm && 0.425 mm\\
  $\sigma_z(0)$ & 0.2 mm  & & 0.425 mm\\
  $\gamma_{x}(0)$ & & 0.0193 $\mu$s$^{-1}$ \\
  $\gamma_{y}(0)$ & & 0.0193 $\mu$s$^{-1}$ \\
   $T_{x}(0)$ & & 0.005 K \\
  $T_{y}(0)$ & & 0.005 K \\
  $T_{z}(0)$ & & 0.6 K \\
 $\rho_{NO}(0)$ & & $1.6 \times 10^{14}$ cm$^{-3}$ \\
 \hline
\end{tabular}
\vspace{-15pt}
\end{table}

\subsection{Formation of a molecular Rydberg gas and its evolution to ultracold plasma}

\begin{figure}
\vspace{-5 pt}
\centering
\includegraphics[width= .5 \textwidth]{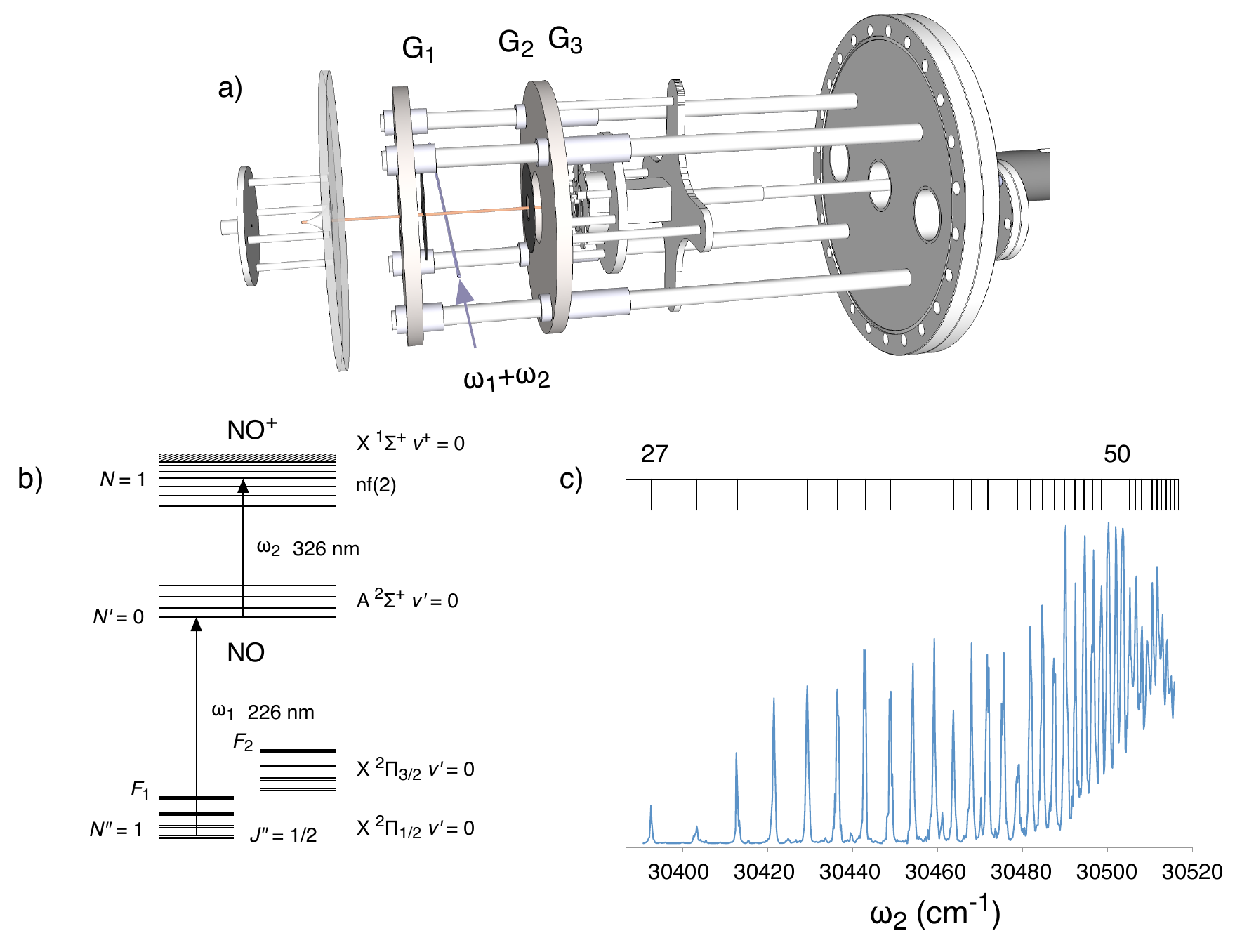}
   \caption{a) Moving grid time-of-flight molecular beam plasma spectrometer.  Co-propagating laser beams, $\omega_1$ and $\omega_2$, cross the molecular beam between entrance aperture, G$_1$, and grid, G$_2$.  b) Level diagram for double-resonant preparation of a state-selected $n_0f(2)$ Rydberg gas.  c) Excitation spectrum of late-peak intensity as a function of $\omega_2$ frequency, structure assigned to the $nf(2)$ Rydberg series. }
\label{fig:apparatus}
\end{figure}

A pair of Nd:YAG pumped dye laser pulses ($\omega_1 + \omega_2$) excite nitric oxide to form a state-selected Rydberg gas.  Figure \ref{fig:apparatus} diagrams a variable flight path apparatus used for short-time experiments \cite{Rennick2011,jphysb}.  Co-propagating laser pulses, $\omega_1$ and $\omega_2$, cross the molecular beam between an entrance aperture, G$_1$, and grid, G$_2$.  

Tuning $\omega_1$ precisely to resonance with the X $^2\Pi_{1/2}$ $N'' = 1$ to A $^2\Sigma^+$ $N'=0$ transition assures that absorption of $\omega_2$ strictly populates Rydberg states with total angular momentum neglecting spin, $N=1$.  Among accessible excited states, only those in the $f$ series converging to NO$^+$, X $^1\Sigma^+$, $N^+=2$ have sufficient lifetime to form a Rydberg gas that can evolve to plasma.  

The pulse energy of $\omega_1$ regulates the density formed at a selected transition up to a maximum of $6 \times 10^{12}$ cm$^{-3}$, obtained by saturating $\omega_1$ and $\omega_2$ \cite{tutorial}.  Varying $\omega_1$ over the linear range from 1 to 6 $\mu$J, produces Rydberg gas distributions with peak density from $1 \times 10^{11}$ to $1 \times 10^{12}$ cm$^{-3}$ \cite{MSW_bifur}.  For a fixed $\omega_1$ intensity, the radiative lifetime of the A $^2\Sigma^+$ state provides a precise means of varying the initial Rydberg gas density by choosing the $\omega_1 - \omega_2$ delay.  

We detect a spectrum of $\omega_2$ transitions that form state-selected $n_0f(2)$ Rydberg gasses in either of two ways.  A field ramped from 0 to 800 V cm$^{-1}$ with a rise-time of 1 $\mu$s, begun at a chosen time after the pair of laser pulses, yields an electron signal by selective field ionization (SFI) \cite{Hung2014}.  Here, the signal, recorded as a function of the ramp voltage forms a spectrum of electron binding energies as the Rydberg gas evolves to plasma.  

Or, in the absence of a field, the volume of excited gas travels with the laboratory velocity of the molecular beam to transit grid, G$_2$, where it encounters a static potential applied to G$_3$, set typically to a value between 30 and 100 V cm$^{-1}$.  The electron signal produced as the excited volume transits G$_2$ forms a late peak that traces the width of the evolving plasma in the $z$ coordinate direction of propagation.  

\begin{figure}
\includegraphics[width= .5 \textwidth]{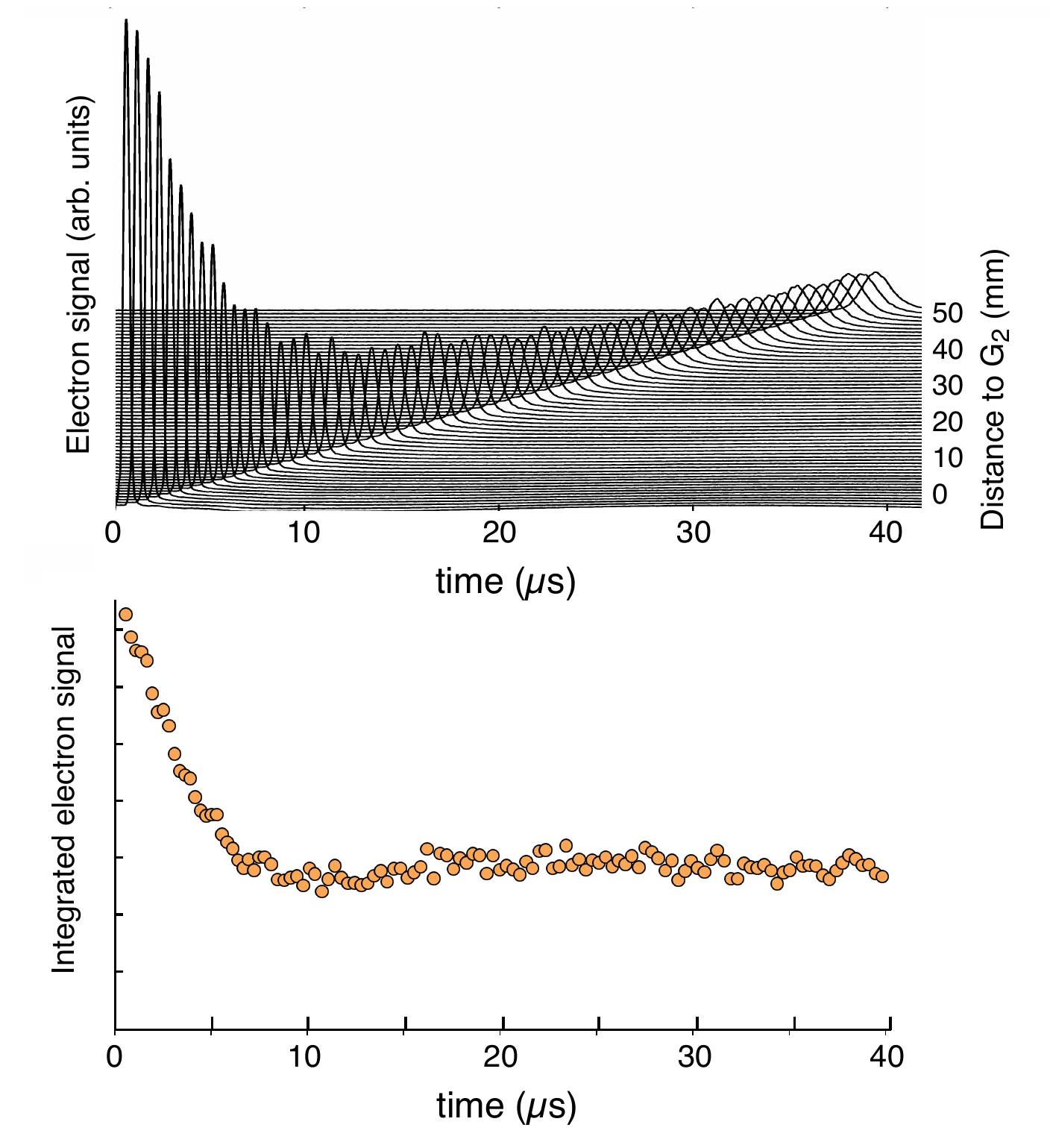}
   \caption{upper frame) Set of traces showing the charge density and width in $z$ of a plasma formed by avalanche from a $50f(2)$ Rydberg gas as a function of flight time for a sequence of flight distances to grid G$_2$, in the apparatus pictured in Figure \ref{fig:apparatus}. (lower frame) Integrated plasma signal obtained as the areas of the traces in the upper frame, plotted as a function of arrival time at G$_2$.}
   \label{fig:traces}
\end{figure}

Waveforms collected as a function of the longitudinal displacement of the detector carriage tell us the rate at which the gas volume expands in the $z$ dimension.  Figure \ref{fig:apparatus} shows the $n_0f(2)$ excitation spectrum of a Rydberg gas sampled by the late-peak signal at G$_2$.  Figure \ref{fig:traces} plots the evolution of width observed by systematically displacing G$_2$ for an $\omega_1$ pulse energy of 1.75 $\mu$J,.  Here we see that the illuminated volume expands very slowly in the direction of propagation.  The width measured in $z$ grows no faster than twice the free expansion rate of a marked volume of gas with the 0.6 K longitudinal temperature of the molecular beam. 

Nitric oxide Rydberg states in the $n_0f(2)$ series predissociate with rates that vary as a function of principal quantum number \cite{Bixon}.  Models predict a general trend in which decay rates fall with increasing $n$ as $n^{-3}$.  Orbital angular momentum also affects this rate \cite{GallagherNO,Remacle}, and the excitation spectrum in Figure \ref{fig:apparatus} shows modulations where interloping Rydberg states of lower $\ell$ form complex resonances of decreased intensity at various points in the $nf$ series.  

Late-peak traces in Figure \ref{fig:traces} show time-dependent signs of neutral dissociation in the plasma system produced by a Rydberg gas with the initial state, $50f(2)$.  In this case, a Rydberg gas  evolves to plasma with a density of $\sim 3 \times 10^{11}$ cm$^{-3}$ on a timescale of a few hundred nanoseconds \cite{MSW_bifur}.  The evident decay of this signal with flight time to G$_2$ reflects the loss of NO$^+$ via its quasi-equilibrium with predissociating NO Rydberg molecules and its direct dissociative recombination with electrons to form N + O atoms \cite{PCCP,Saquet}.  

Over the first 10 $\mu$s, neutral dissociation processes combined with expansion reduce electron temperature and the ion-electron density.  Thereafter, plasma decay apparently ceases, yielding an integrated signal of constant area.  Extended measurements using Ar as a molecular beam carrier gas confirm that this signal persists undiminished for evolution times as long as 180 $\mu$s \cite{Haenel_width}.

As detailed below, our imaging apparatus measures the widths of observed plasma volumes as a function of time in all three Cartesian dimensions \cite{MSW_bifur}.  On the basis of these measurements, we can estimate that the $50f(2)$ Rydberg gas ellipsoid with initial Gaussian dimensions, $\sigma_x=0.75$, $\sigma_y=0.42$, $\sigma_z=0.42$ mm forms a plasma that expands in 10 $\mu$s to shape defined approximately by Gaussian ellipsoid widths of, $\sigma_x=1.0$, $\sigma_y=0.55$, $\sigma_z=0.70$ mm.  Combining this observed change in volume with the evident decrease of integrated signal in Figure \ref{fig:traces} tells us that the plasma formed with an initial peak density of $3 \times 10^{11}$ cm$^{-3}$ decays and expands after 10 $\mu$s to a peak density of $4 \times 10^{10}~{\rm cm^{-3}}$. 

Electron signal measurements in our longer flight path imaging instrument yield strong waveforms of equal area after flight times of 200 and 400 $\mu$s (see below), suggesting a plasma lifetime substantially in excess of 1 millisecond.  Accounting for the measured expansion in $x, y$ and $z$, we can estimate that after a flight time of 400 $\mu$s, a plasma described by the waveforms in Figure \ref{fig:traces} strikes our imaging detector with a peak density on the order of $1 \times 10^{7}~{\rm cm^{-3}}$.

\subsection{Selective field-ionization spectrum as a probe of the relaxation from Rydberg gas to plasma}

A Rydberg gas in a quantum state $n$ undergoes selective field ionization (SFI) when the amplitude of a ramped electrostatic field reaches the electron binding energy threshold for that state \cite{RydbergAtoms}.  The free electrons bound by the space charge of a low-density plasma separate from the ions at an early point in a typical field ramp \cite{WalzFlannigan}, affording a field ionization trace that differs very little from the SFI spectrum of a Rydberg gas of very high principal quantum number.

\subsubsection{Absolute calibration of the initial density of the Rydberg gas in a molecular beam}

\begin{figure*}
\centering
\includegraphics[width= 1 \textwidth]{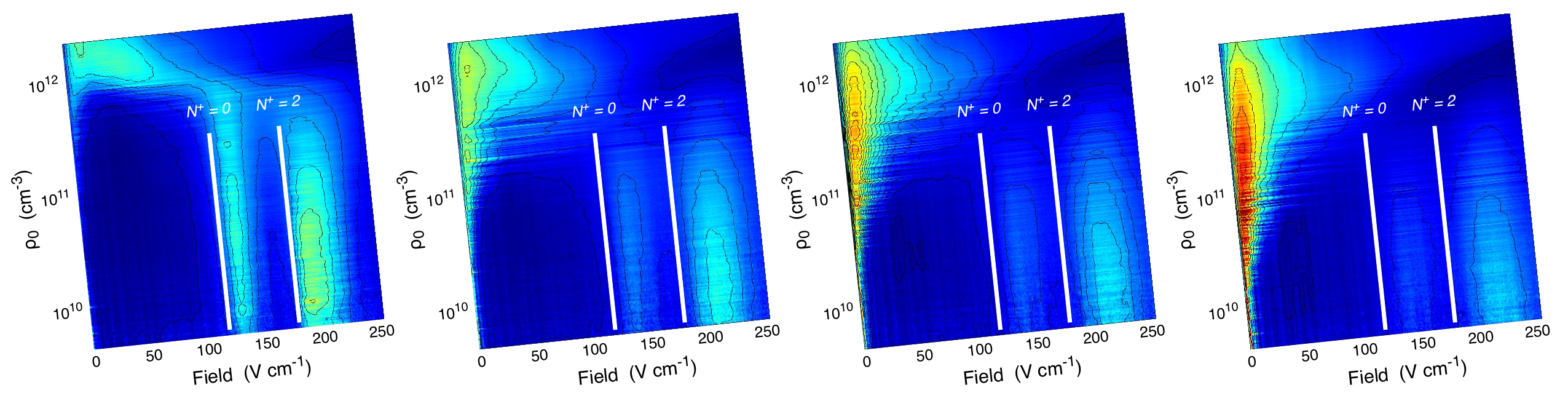}
   \caption{\small (left) \textbf{Selective field ionization spectra of NO:} Contour plots showing SFI signal as a function the applied field for an $nf(2)$ Rydberg gas with an initial principal quantum number, $n_0=44$.  Each frame represents 4,000 SFI traces, sorted by initial Rydberg gas density.  Ramp field potential applied to G$_1$, beginning from left to right 0, 150, 300 and 450 ns after the $\omega_2$ laser pulse.  The two bars of signal most evident at early ramp field delay times represent the field ionization of the $44f(2)$ Rydberg state respectively to NO$^+$ X $^1\Sigma^+$ cation rotational states, $N^+=0$ and 2.  This signal shifts to higher field with increasing ramp-field delay owing to $\ell$-mixing \cite{WalzFlannigan}.  The signal waveform extracted near zero applied field represents the growing population of plasma electrons.  Refer to Figure \ref{fig:Vlasov} for a false-colour scale bar.
   }
\label{fig:SFI}
\end{figure*}

The total charge collected in the pulsed-field ionization of a given volume of Rydberg gas at any stage in its evolution to plasma relates directly to its initial excited state density.  We have confirmed this in experiments on Rydberg gases of nitric oxide, examining the total yield of electrons observed as a function of $\omega_1$ pulse energy, from 1 to 20 $\mu$J, and $\omega_1$ - $\omega_2$ delay ($\Delta t_{\omega_2}$) from 0 to 150 ns.  

For low laser intensities in this range, $\omega_1$ produces a density of molecules excited to the A $^2\Sigma ^+$ intermediate state that varies linearly with laser pulse energy, while laser pulses with energies above 15 $\mu$J clearly saturate this first excitation step.  

A precise knowledge of the molecular beam density enables us to specify the maximum density of A-state NO under conditions of saturated $\omega_1$ excitation.  The well-known radiative lifetime of the NO A $^2\Sigma ^+$ state precisely determines the relative excited state density as a function of $\Delta t_{\omega_2}$ for any $\omega_1$ power \cite{Crosley}.  

We use these two well-defined properties:  i) the approach to saturated excitation of the X to A transition, and ii) the decay of A-state population in time, to build a robust classification model for Rydberg gas density.  This model uses the shot-to-shot integrated electron signal to determine the absolute initial Rydberg gas density for every shot in any family of SFI experiments that includes some $\omega_1$ laser pulse energies in the saturated regime.

\subsubsection{SFI as a probe of avalanche dynamics}

The double-resonant excitation of NO forms an ellipsoidal volume of Rydberg gas with a well-defined density distribution.  The absolute magnitude of the density in each element of this volume determines the fraction of molecules with nearest-neighbour distances that fall within a critical radius for prompt Penning ionization.  Trapped by the NO$^+$ space charge, Penning electrons collide inelastically with Rydberg molecules.  Relaxation releases energy, and the system builds to an electron-impact avalanche.   

Classifying individual SFI traces for initial Rydberg gas density provides a revealing gauge of the avalanche dynamics.  Sequences of frames in Figure \ref{fig:SFI} show families of 4,000 SFI traces for an initial principal quantum number of $n_0=44$, sorted in each frame by initial Rydberg gas density from 10$^{12}$  to 10$^{10}$ cm$^{-3}$.  We have recorded many such SFI traces for values of $n_0$ from 30 to 60.  These measurements detail the evolution of a molecular nitric oxide Rydberg gas to plasma as a function of its calibrated density and selected initial principal quantum number.  

Here, for densities greater than 10$^{12}$ cm$^{-3}$, we see that the field ionization structure of the $n_0f(2)$ Rydberg state gives way to a low-field extraction of plasma electrons on a time scale faster than the 1 V ns$^{-1}$ rise time of the field ramp.  At two orders of magnitude lower Rydberg gas density, very little free electron signal develops on a half-microsecond time scale.  

We and others have shown that simple coupled rate-equation calculations account well for the avalanche dynamics of atomic systems, as modelled by MD simulations, provided that these treatments define consistent limits on the depth of Rydberg binding energy \cite{Niffenegger,Morrison2012}.  

Rydberg molecules formed in the molecular NO plasma predissociate, and we represent the avalanche in this case with an extended set of coupled differential equations \cite{Saquet}:  

\begin{eqnarray}
-\frac{d\rho_i}{dt}&=&\sum_{j}{k_{ij}\rho_e\rho_i}-\sum_{j}{k_{ji}\rho_e\rho_j} \nonumber\\
&& +k_{i,ion}\rho_e\rho_i-k_{i,tbr}\rho^3_e + k_{i,PD}\rho_i
  \label{level_i}
\end{eqnarray}
\noindent and,
\begin{equation}
\frac{d\rho_e}{dt}=\sum_{i}{k_{ion}\rho_e^2}-\sum_{i}{k^i_{tbr}\rho^3_e} - k_{DR}\rho^2_e
  \label{electron}
\end{equation}
\noindent in which a variational reaction rate formalism determines $T_e$-dependant rate coefficients, $k_{ij}$, for electron impact transitions from Rydberg state $i$ to $j$, $k^i_{ion}$, for collisional ionization from state $i$ and $k^i_{tbr}$, for three-body recombination to state $i$ \cite{Mansbach,PVS}.  Unimolecular rate constant, $k_{i,PD}$, describes the principal quantum number dependant rate of Rydberg predissociation \cite{Bixon,GallagherNO,Remacle}, averaged over azimuthal quantum number, $l$ \cite{Chupka:1993}.  $k_{DR}$ accounts for direct dissociative recombination \cite{Schneider}

The temperature of electrons released by avalanche balances with the relaxation of molecules in the manifold of  Rydberg states populated by initial Penning ionization and subsequent three-body recombination, conserving total energy per unit volume:
\begin{eqnarray}
E_{tot}&=&\frac{3}{2}k_BT_e(t)\rho_e(t)-R\sum_i{\frac{\rho_i(t)}{n_i^2}} \nonumber \\
&&+ \frac{3}{2}k_B T \rho_e^{DR} -R \sum_{i} \frac{\rho_i^{PD}}{n_i^2}
  \label{energy}
\end{eqnarray}
where $R$ is the Rydberg constant for NO, and $\rho_e^{DR}$ and $\rho_i^{PD}$ represent the number of electrons and Rydberg molecules of level $i$ lost to dissociative recombination and predissociation, respectively.

For present purposes, we consider neutral plasmas of uniform ion and electron density, denoted above as $\rho_e$.  Short timescale simulations neglect the predissociation of nitric oxide Rydberg molecules and dissociative recombination of NO$^+$ ions..

\begin{figure}[h!]
\centering
\includegraphics[width= .48 \textwidth]{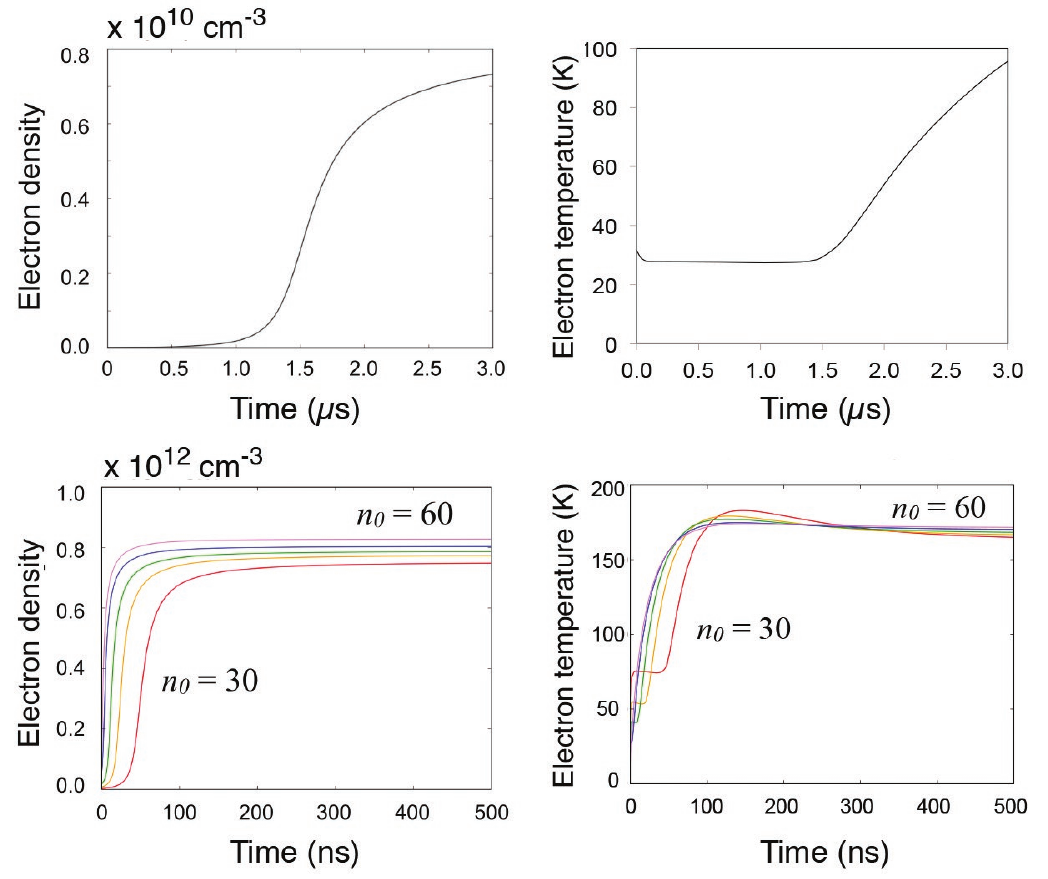} 
   \caption{Rate equation simulations of the electron-impact avalanche of a Rydberg gas: (upper curves) Predicted growth in electron density and temperature on a microsecond timescale for a $n_0=50$ Rydberg gas with initial density $\rho=1 \times 10^{10}$ cm$^{-3}$.  (lower curves) Change in electron density and temperature on a nanosecond timescale for Rydberg gases with principal quantum numbers $n_0=30, 35, 40, 50$ and 60 with an initial density $\rho=1 \times 10^{12}$ cm$^{-3}$.  }
\label{fig:rate_eqn}
\end{figure}
Figure \ref{fig:rate_eqn} shows the growth of electron density and electron temperature determined by rate-equation models for avalanche in Rydberg systems with uniform initial densities of 10$^{10}$ and 10$^{12}$ cm$^{-3}$.  Generally, rate equations predict weak variations in rise time and quasi-equilibrium electron temperature with initial principal quantum number, but these avalanche characteristics depend sensitively on initial Rydberg gas density.  In model calculations, lower-density Rydberg gases evolve slowly to form plasmas with electron temperatures in the range of 100 K.  Simulated avalanches at a density of 10$^{12}$ proceed on a nanosecond timescale and form electron gases with temperatures near $T_e = 180$ K.   

Longer-timescale coupled rate-equation simulations for molecular plasmas with Gaussian ellipsoidal density distributions differ in important ways, as detailed below.  Nevertheless, regardless of density gradients and dissociation, electron collisions that relax Rydberg molecules and increase $T_e$ figure prominently in the the short time dynamics of electron impact avalanche in a Rydberg gas.  

\subsection{Plasma expansion as a measure of electron temperature}

\begin{figure*}
\centering
\includegraphics[width= 0.9 \textwidth]{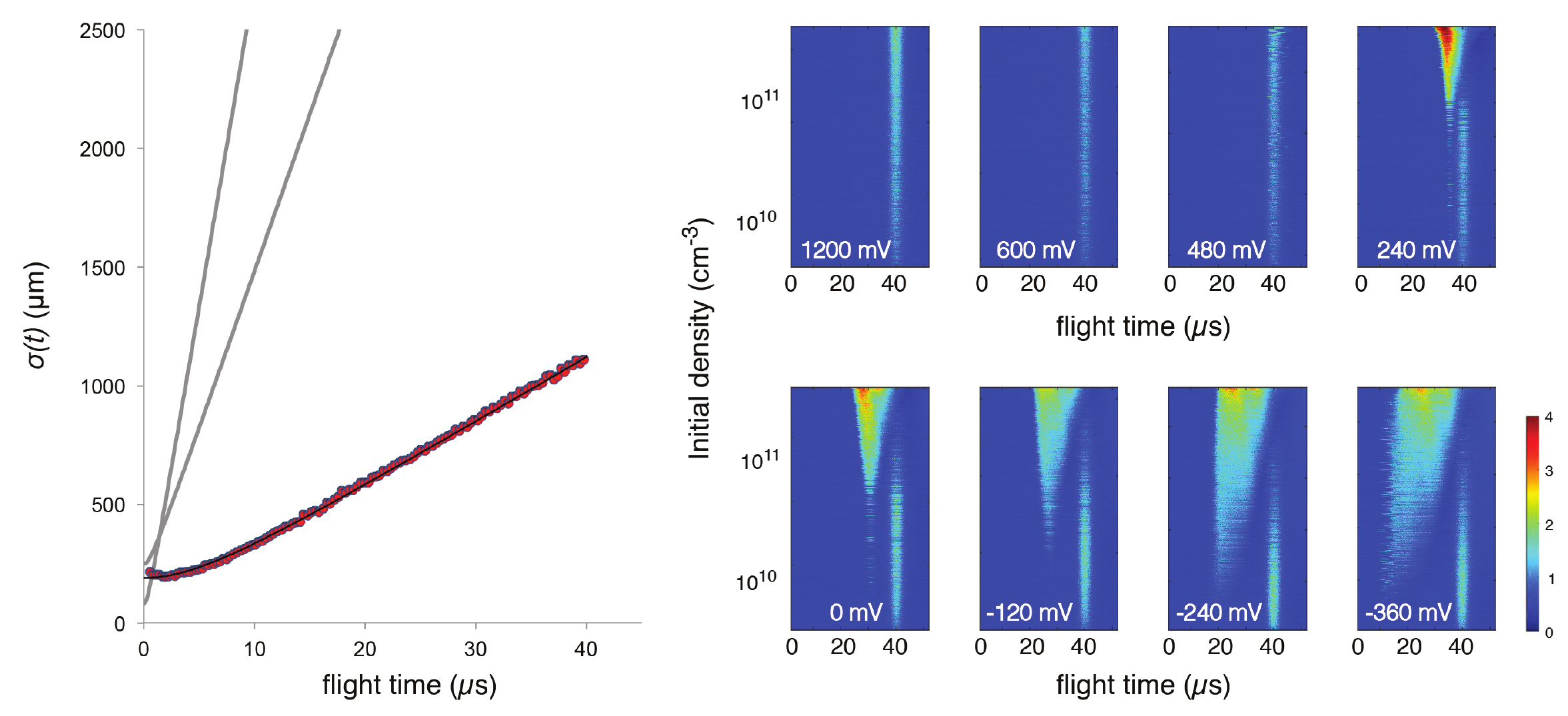} 
   \caption{\small (left) Self-similar expansion of Gaussian ultracold plasmas: (grey lines) Ambipolar expansion of a model Gaussian plasma core ellipsoid of cold ions and $T_e=180$ K electrons with $\sigma_y(0) = \sigma_z(0) = 83~\mu$m and $\sigma_x(0) = 250~\mu$m. Ions rapidly attain ballistic velocities, $\partial_t \sigma_y = \partial_t \sigma_z = 272$ m s$^{-1}$ and $\partial_t \sigma_x = 132$ m s$^{-1}$.  (blue line with red data points) Experimental measure of $\partial_t \sigma_z(t)$ fit by Vlasov model for a Gaussian spherical expansion for $T_e= 5$ K.  (right) Electron signal as a function of flight time to G$_2$ and Rydberg gas principal quantum number observed in the moving grid apparatus diagrammed in Figure \ref{fig:apparatus} with a constant reverse bias on G$_1$ of 1.20, 0.60, 0.48 and 0.24 V (top row), and forward bias of 0.0, $-0.12$, $-0.24$ and $-0.36$ V, all with a flight distance to G$_2$ of 56 mm.}
\label{fig:Vlasov}
\end{figure*}

The relaxation of Rydberg molecules in the ionization avalanche dramatically increases the temperature of plasma electrons.  In a freely expanding plasma, the growing ambipolar pressure of this electron gas radially accelerates the ions.  The Vlasov equations describe this acceleration analytically for the self-similar expansion of a spherical Gaussian plasma \cite{Laha,Killian_Phys_rept}.  We and others have extended these hydrodynamics both analytically and numerically to non-spherical plasma distributions \cite{Sadeghi:2012,Cummings,tutorial}.  

In the limit of saturated $\omega_1$ excitation, laser-crossed molecular beam illumination creates an ellipsoidal volume of Rydberg gas with a peak density of $\sim 10^{12}$ cm$^{-3}$.  A coupled rate-equation model (c.f. Figure \ref{fig:rate_eqn}) for a regime of this density predicts avalanche on a nanosecond timescale to form an electron gas with an initial temperature, $T_e(0)=180$ K.  The pressure of this electron gas drives the expansion of the ions.  The steeper charge-density gradient in the $y,z$ coordinate directions causes the rate of radial expansion in the short axis plane to substantially exceed that along the cross-beam $x$ axis of the ellipsoid.

Grey curves in Figure \ref{fig:Vlasov} diagram the self-similar ambipolar expansion of Gaussian plasma ellipsoid of cold ions and $T_e=180$ K electrons in a subspace with $\sigma_y(0) = \sigma_z(0) = 83~\mu$m and $\sigma_x(0) = 250~\mu$m.  In a Vlasov model, electrons of this temperature accelerate the distribution of NO$^+$ ions to a $z$ expansion rate, $\partial_t \sigma_z$, of  272 m s$^{-1}$.  

This predicted radial velocity, which approaches 20\% of the laboratory speed of the molecular beam, greatly exceeds the observed expansion rate of the ultracold plasma in $z$ determined experimentally by the growing widths of waveforms such as those displayed in Figure \ref{fig:traces}.  A typical set of late-peak Gaussian widths yields values of $\sigma_z(t)$ plotted as the red data points in Figure \ref{fig:Vlasov}.

The right frame of Figure \ref{fig:Vlasov} provides clear evidence for the presence of two distinct domains of ultracold plasma evolution.  Here a forward bias between G$_1$ and G$_2$ reveals evidence for an avalanche of a high-density Rydberg gas that produces a diffuse, rapidly expanding plasma of energetic ions and electrons.  A reverse bias of 480 mV cm$^{-1}$ or more sweeps these electrons away, isolating the slowly expanding waveform of the ultracold plasma.  But, the signal of electrons in advance of the late peak waveform persists up to a reverse bias as great as 250 mV cm$^{-1}$, signifying an initial fast-component $T_e$ of 200 K.

Two questions naturally arise:  What aspect of the Rydberg gas relaxation dynamics distinguishes these two processes, which separate so evidently in the molecular beam propagation direction, $z$?  How do these dynamics lead to the formation of a strongly coupled ultracold plasma?  To answer these questions, we turn to experiments that image the hydrodynamics of plasma expansion in the $x,y$-plane.

\subsection{Ultracold plasma hydrodynamics in three dimensions}

\subsubsection{Plasma evolution in a Rydberg gas of non-uniform density} \label{evo}

Laser-crossed molecular beam excitation forms an ellipsoidal Rydberg gas in which density distributes differently in $x$, $y$ and $z$.  The local density of ground-state NO does not vary in the molecular beam propagation direction, $z$, and thus the Rydberg gas density distribution in this coordinate simply reflects the narrow Gaussian intensity profile of the laser beam.  The Rydberg gas has the broader Gaussian width of the molecular beam in the $x$ coordinate direction, defined by the propagation of the laser.  The profiles of the laser beam and molecular beam combine to form a steeper gradient of Rydberg gas density in $y$, determined by the product of these Gaussians.  

The NO$^*$ Rydberg molecules in the volume marked by $\omega_1$ have initial velocity distributions defined by temperatures, $T_k(0)$ and divergence, $\gamma_k(0)$, where $k$ refers to the coordinate directions, $x$, $y$ and $z$.  As indicated in Table 1, molecules in the skimmed supersonic beam have a temperature of 0.6 K in the longitudinal direction with no divergence.  Initial velocity distributions in $x$ and $y$ are narrower, with $T_\perp = 5$ mK and radial velocity, $\gamma_k r_k$, where $r_k$ refers to radial displacement in the $k$ direction and $\gamma_k = 0.0193~\mu$s$^{-1}$ for $k= x$ and $y$.

The superimposed distributions of laser intensity and molecular beam number density confine the peak density of NO$^*$ Rydberg molecules to a smaller ellipsoid in the core of the Rydberg gas.  There, rate-equation simulations predict avalanche on a nanosecond timescale, forming a local population of free electrons in a quasi equilibrium with initial temperature $T_e(0)$ as high as 180 K.  Such a high electron temperature must drive plasma expansion.  

For the purposes of a conceptual model, let us represent the density distribution of this core plasma by means of a set of concentric ellipsoidal shells.  To simplify, we neglect both the thermal motion of the ions and any process that leads to neutralization, such as dissociative or three-body recombination, so that the number of charged particles in each shell remains constant.  

In the quasi-neutral  approximation, an electric potential gradient gives rise to a force, $-e\nabla \phi_{k,j}(t)$, that accelerates the ions in shell $j$ in direction $k$ according to \cite{Sadeghi:2012}:
\begin{align}
\frac{-e}{m'}\nabla \phi_{k,j}(t) = & \frac{\partial u_{k,j}(t)}{\partial t} \notag  \\
= & \frac{k_BT_e(t)}{m'\rho_j(t)} \frac{\rho_{j+1}(t) - \rho_j(t)}{r_{k,j+1}(t) - r_{k,j}(t)}
  \label{dr_dt}
\end{align}
\noindent where $\rho_j(t)$ represents the density of ions in shell $j$.  

The instantaneous velocity, $u_{k,j}(t)$ determines the change in the radial coordinates of each shell, $r_{k,j}(t)$, 
\begin{equation}
\frac{\partial r_{k,j}(t)}{\partial t}=u_{k,j}(t) = \gamma_{k,j}(t) r_{k,j}(t)
  \label{dr_dt}
\end{equation}
\noindent which in turn determines shell volume and thus its density, $ \rho_j(t)$.  
\begin{figure*}[t]
\centering
\includegraphics[width= .9 \textwidth]{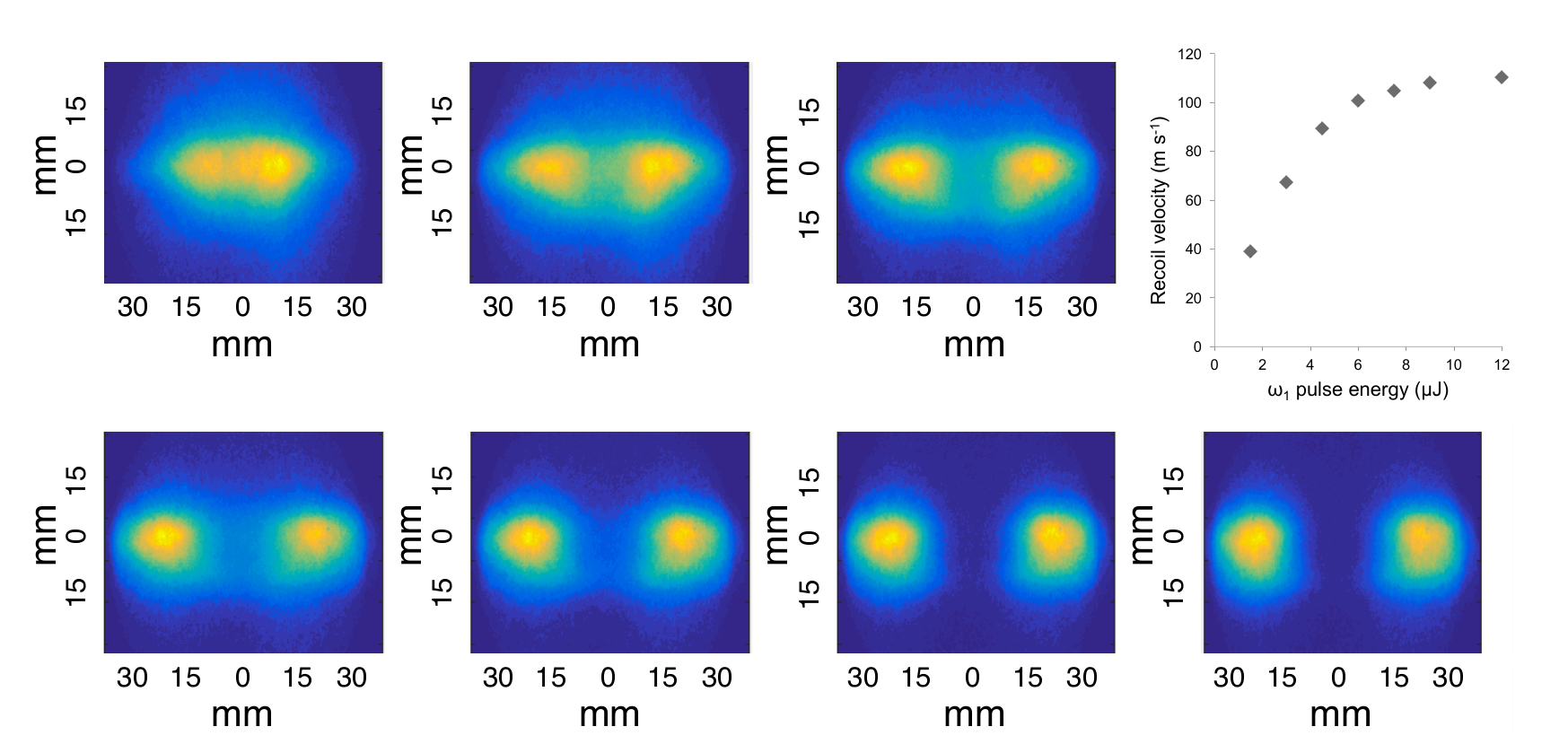}  
   \caption{\small Plasma bifurcation and recoil:  $x,y$ detector images of ultracold plasma volumes produced by 2:1 aspect ratio ellipsoidal Rydberg gases with selected initial state, $40f(2)$ after a flight time of 402 $\mu$s over a distance of 575 mm.  Charge distributions bifurcate with increasing recoil velocity as a function of $\omega_1$ laser pulse energies of 1.5, 3, 4.5, 6, 7.5, 9 and 12 $\mu$J.  Inset: Measured recoil velocity plotted as a function of $\omega_1$ laser power.  Refer to Figure \ref{fig:Vlasov} for a false-colour scale bar.}
\label{image1}
\end{figure*}
The electron temperature supplies the thermal energy that drives this ambipolar expansion.  Ions accelerate and $T_e$ falls according to: 
\begin{equation}
\frac{3k_B}{2}\frac{\partial T_e(t)}{\partial t}= -\frac{m'}{\sum_{j}{N_j}}\sum_{k,j}{N_j u_{k,j}(t)\frac{\partial u_{k,j}(t)}{\partial t}}
  \label{dr_dt}
\end{equation}
\noindent where we have defined an effective ion mass that recognizes the redistribution of ion momentum by free-electron-mediated resonant ion-Rydberg charge exchange, which occurs with a very large cross section \cite{PPR}.  
\begin{equation}
m' =\left (1+ \frac{\rho^*_{j}(t)}{ \rho_j(t)}\right) m
  \label{dr_dt}
\end{equation}
\noindent in which $\rho^*_{j}(t)$ represents the instantaneous Rydberg density in shell $j$.

The initial avalanche in the high-density core of the ellipsoid leaves few Rydberg molecules, so this term has little initial effect.  Rydberg molecules predominate in the lower-density wings.  There momentum sharing by charge exchange assumes greater importance.  

As indicated by the high-temperature expansion curves in Figure \ref{fig:Vlasov}, core ions reach ballistic velocities in just a few hundred nanoseconds.  This quenches electron temperature.  After expanding less than 50 $\mu$m, the motion of the core ellipsoid becomes a flux of ballistic ions and very cold electrons.  

Much of this population streams out of the Rydberg gas volume in the $y$ and $z$ directions to produce the early signal pictured in Figure \ref{fig:Vlasov}.  But, a substantial fraction of these expanding ions and electrons remain in the volume of Rydberg gas, streaming in $\pm x$.  Here, ion-Rydberg charge exchange has two important consequences:  

(1) A rapid sequence of electron transfer processes acts to redistribute the directed momentum of the ions over the entire population of Rydberg molecules and ions \cite{Becker}.  The heavy particles relax to correlated positions \cite{Sadeghi:2014}, and the two volumes stream as a whole in opposite directions.  

(2) Charge exchange increases the effective inertial mass of the ions, which acts to retard the ambipolar expansion of the plasma forming in the bifurcated volumes.  

\subsubsection{Molecular beam ultracold plasma imaging spectrometer} 

Electron signal waveforms of the charge density distribution in $z$ provide ample evidence for the production of an initial hot electron plasma, accompanied by a cold component that appears to expand slowly in the axis of molecular beam propagation.  The previous section describes a sequence of hydrodynamic processes that convert the thermal energy of hot electrons to the streaming velocities of opposing volumes of cold ions and entrained Rydberg molecules.  We look now for experimental evidence of these effects in the electron density distributions captured by the molecular beam ultracold plasma imaging spectrometer.  

This experiment uses the apparatus diagrammed in Figure \ref{fig:imaging}.  Here, a skimmed molecular beam enters a field-free flight tube capped by a multichannel plate detector with a phosphor-screen anode.  Co-propagating laser pulses, $\omega_1+\omega_2$ cross the molecular beam to form a Rydberg gas 75 mm beyond the skimmer marking a 2:1 ellipsoidal volume with the hydrodynamic properties listed in Table 1.  The excited volume travels a selected distance of 325 or 600 mm to strike the detector, which images the $x,y$ distribution of electrons, integrated in $z$.  

This apparatus yields images that clearly show the bifurcation predicted above.  Opposing volumes, formed by evolution from Rydberg gas to ultracold plasma separate with a $\pm x$ recoil velocity that depends systematically on the initial density of the Rydberg gas, $\rho_0$, and its selected principal quantum number, $n_0$ \cite{MSW_bifur}.  

Figure \ref{image1} shows a sequence of seven images recorded after a flight path of 575 mm for a 2:1 ellipsoidal Rydberg gas prepared in $n_0=40$ with $\omega_1$ pulse energies from 1.5 to 12 $\mu$J.  Note how the the recoil velocity varies with $\omega_1$ power, apparently saturating as the laser pulse energy reaches 8 $\mu$J.  Fits to the electron-signal images at higher pulse energy consistently yield Gaussian widths, $\sigma_x = \sigma_y = 6.4$ mm.  

\begin{figure}
\centering
\includegraphics[width= .5 \textwidth]{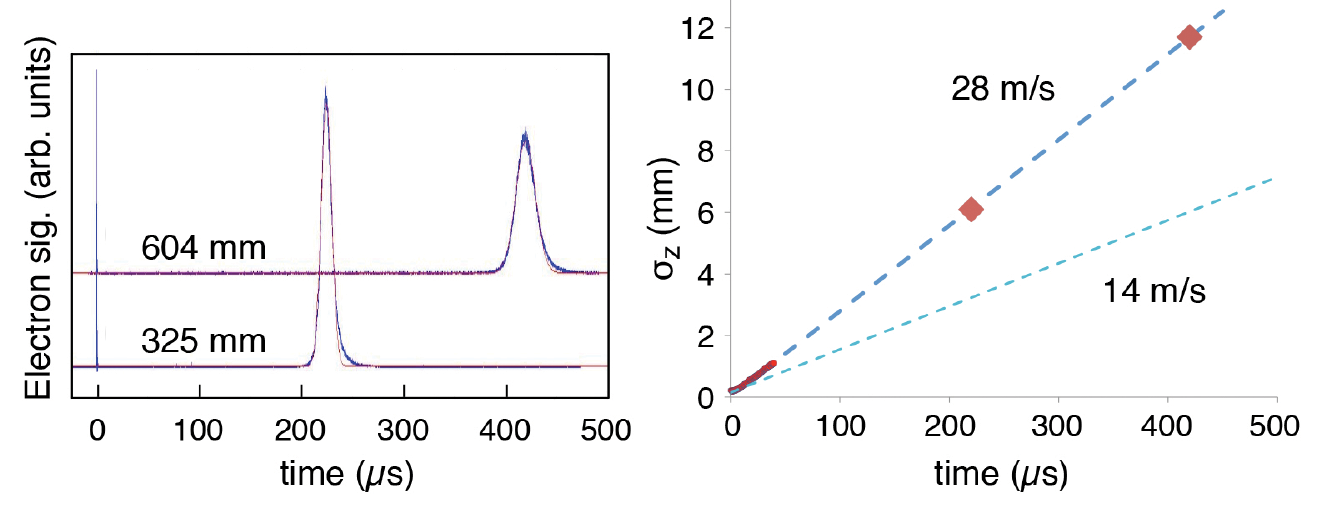}  
 \caption{\small  Long-time dynamics of recoiling plasma volumes:  (left) electron signal waveforms in $z$ obtained for $n_0 = 65$ at short and long flight distances.  (right) Expansion measured by $\sigma_z(t)$ compared with the short-time expansion observed in the moving grid spectrometer, and the 14 m s$^{-1}$ expansion in $z$ of the ellipsoidal volume marked by $\omega_1$ }
\label{image2}
\end{figure}

The electron-signal waveform obtained by collecting the anode current as a function of time gauges the width of this plasma in the $z$ coordinate.  Figure \ref{image2} shows waveforms obtained in the molecular beam plasma imaging spectrometer for flight distances of 325 and 604 mm.  As indicated by the plot of $\sigma_z$ versus time, these widths - measured after exceptionally long flight times - conform precisely with the very slow rates of expansion, determined for short times of flight in our moving grid machine (c.f. Figures \ref{fig:traces} and \ref{fig:Vlasov}).  After 400 $\mu$s, the plasma $x,y$ distributions pictured in Figure \ref{image1} have a Gaussian width in z of $\sim12$ mm.  

The slow expansion rate of this enduring plasma component over its entire trajectory puts a significant limit on the kinetic energy of its free electrons.  Shell models in the geometry of the plasma ellipsoid produce expansions this slow only for electron temperatures no more than a few degrees Kelvin \cite{tutorial}.  

Avalanche and fast expansion in the high-density core of the Rydberg gas ellipsoid burns a hole that adds to the apparent recoil velocity of the plasma volumes in Figure \ref{image1}.  This hole, which extends to a $\pm x$ width of $1 \times \sigma$ (0.75 mm) at the point of excitation widens by the divergence of the molecular beam, to nearly 3 mm at the detector.  

Interpreted as a recoil velocity, such a structure alone would read as 16 m s$^{-1}$.  Adjusting the estimated saturating recoil velocity for this effect still leaves opposing gas volumes separating at more than 90 m s$^{-1}$.  

In other words, the pair of plasma volumes recoil at nearly half the ballistic velocity of the initial, hot core ions at a radius of $\sigma_x$ in the coordinates of the original ellipsoid.  For this to be possible, the number of ballistic ions and electrons entering the wings of the ellipsoid must compare to number of Rydberg molecules encountered there.  

\begin{figure}[h!]
\centering
\includegraphics[width= .5 \textwidth]{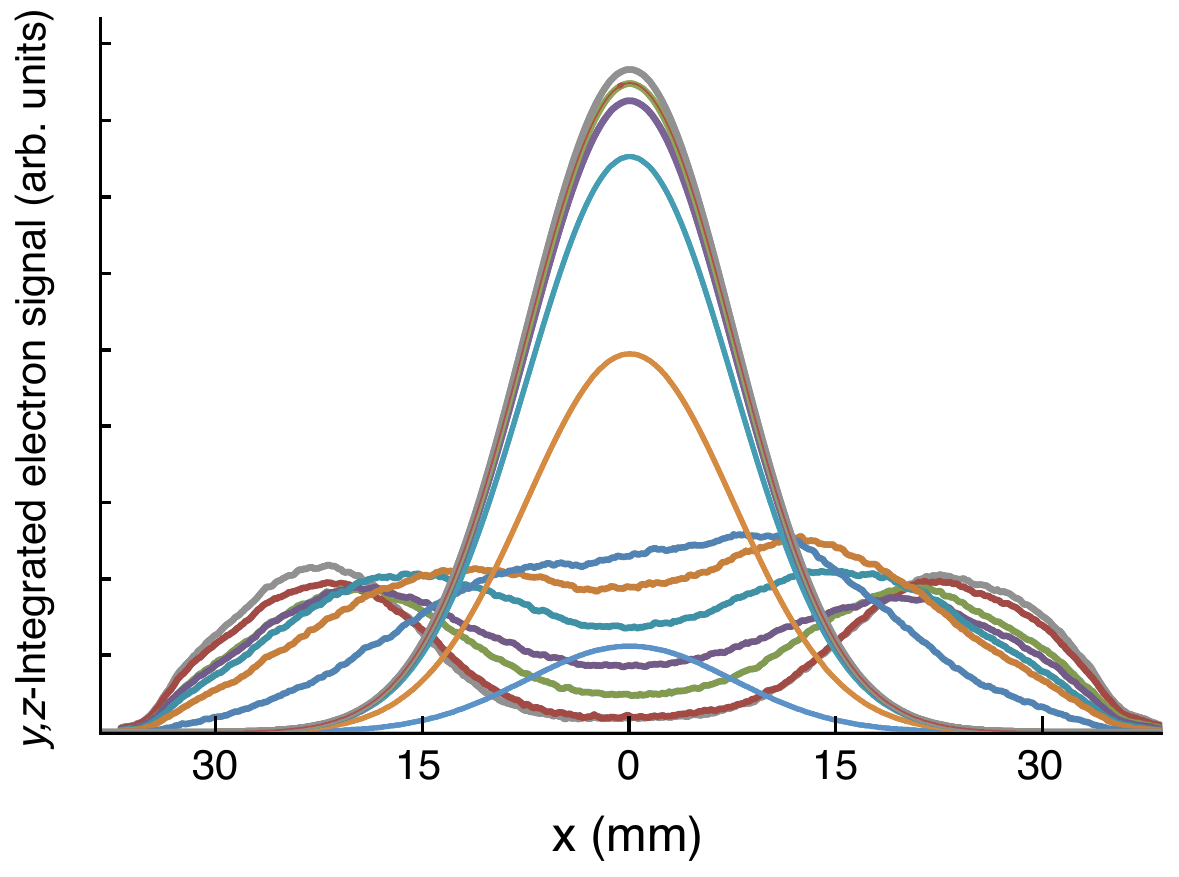} 
   \caption{\small Bifurcated traces giving absolute sums of imaging detector columns in $y$ as a function of $x$ for the images pictured in Figure \ref{image1}, with centred traces integrating to define a lost electron component that accounts for the charge missing in the image traces as the $\omega_1$ power approaches saturation of the X $^2\Pi \rightarrow$ A $^2\Sigma^+$ transition in NO.  Missing charge distributed arbitrarily as a Gaussian with full the $\sigma_x$ of the molecular beam at the point of laser interaction propagated by the beam divergence over the 600 mm flight distance to the detector.}
\label{fig:hole}
\end{figure}

Figure \ref{fig:hole} compares the absolute number distributions of charged particles detected to form the images pictured in Figure \ref{image1}.  Over the $\omega_1$ pulse energy range of this experiment, the original Rydberg gas density grows by a factor of three, approaching saturation in parallel with the increase in recoil velocity pictured in Figure \ref{image1}.  But, in every case, the distributions plotted in Figure \ref{fig:hole} evolve to occupy roughly the {\it same} final volume, and integrate to represent states with virtually the {\it same} number density of charged particles.  

Figure \ref{fig:hole} also includes curves that account for the missing charge density, plotted here as the radial Gaussian of the molecular beam, measured at the point of laser interaction and propagated by the beam divergence over the distance to the detector.  

Figure \ref{fig:traces} traces the number of charged particles as a function of flight time observed for of an ultracold plasma produced in a beam of NO seeded in He.  Here, as described above, an initial plasma density of $3\times 10^{11}$ cm$^{-3}$ falls with the loss of its hot component and dissociative recombination to $4\times 10^{10}$ cm$^{-3}$ after 10 $\mu$s.  Thereafter, the loss of ions and electrons apparently stops, and the plasma exhibits a constant integrated charge for as long as we can let it fly in the moving grid apparatus (180 $\mu$s, seeded in Ar \cite{Haenel_width}).  

The $z$ waveforms in Figure \ref{image2} show that the longer flight-path imaging apparatus produces a signal at 400 $\mu$s with no less an area than the one we observe with the detector positioned for a time of flight of 200 $\mu$s.    The persistent magnitude of these signals suggests that the corresponding NO$^+ -$ electron system lives a millisecond or more.  

The surviving number of NO$^+$ ions and electrons observed at all flight times greater than 10 $\mu$s for an initial Rydberg gas density of $3 \times 10^{11}$ cm$^{-3}$ corresponds to a density of $7 \times 10^{10}$ cm$^{-3}$ in the volume initially defined by the intersection of the $\omega_1$ with the molecular beam.  The plasmas represented by the traces shown in Figures \ref{image1} and \ref{fig:hole} all integrate to about the same total number of charges.  A measurement of the image widths in $x$, $y$ and $z$ establishes that each of these plasmas evolves to occupy approximately the same volume.  Thus, remarkably, we find, after 400 $\mu$s of evolution, that plasmas produced with $\omega_1$ pulse energies from 1.5 to 12 $\mu$J evolve universally to the same, apparently low electron temperature and a common particle density of about 10$^{7}$ cm$^{-3}$.  

In the bifurcated plasma, SFI spectra show that electrons linger at levels near and above the first ionization threshold, despite a potential energy per charge pair almost twice the N-O bond dissociation energy, and a plasma predissociation rate of $6 \times 10^4$ s$^{-1}$ (or lifetime $\tau=17~\mu$s) predicted by coupled rate equation calculations \cite{Saquet}.  The electron signal waveforms displayed in Figures  \ref{image1} and \ref{image2}, as well as all our other long-time observations of nitric oxide molecular ultracold plasma evolution, thus appear to indicate a state of arrested relaxation.

\subsubsection{Summary:  Sequence of events that lead to ultracold plasma bifurcation}

Gaussian laser and molecular beams intersect to create an ellipsoidal Rydberg gas with a selected initial state, $n_0$.  The core avalanches producing a cloud of charged particles that expands with as much as one-fifth the laboratory velocity of the molecular beam, suggesting initial electron temperatures between 100 and 200 K. 

Ambipolar expansion quenches electron kinetic energy.  Accelerated core ions stream in the $\pm x$ direction, into the wings of the Rydberg gas.  There, recurring reactions that exchange charge between NO$^+$ ions and NO$^*$ Rydberg molecules apparently equilibrate velocities and change ion motion within the gas to $\pm x$ motion of gas volumes in the laboratory.  The ion/Rydberg temperature falls, spatial correlation develops, and over a period of 500 ns, the system forms the plasma/high-Rydberg quasi-equilibrium dramatically evidenced by the SFI results in Figure \ref{fig:SFI}.  

In the wings, momentum redistribution owing to continuing charge transfer with the residual high-Rydberg population retards axial expansion \cite{Pohl2003,PPR}.  By redirecting electron energy from ambipolar acceleration to $\pm x$ plasma motion, NO$^+$ to NO$^*$ charge exchange dissipates electron thermal energy, preserves density and enables ions and Rydberg molecules to relax to positions that minimize potential energy and build spatial correlation.    

\subsubsection {Experimental evidence for arrested relaxation } 

Ultracold plasma volumes separate with velocities that scale uniformly with initial Rydberg gas principal quantum number, $n_0$, and density, $\rho_0$ \cite{MSW_bifur}.  The magnitudes of these recoil velocities suggest that the plasma wings incorporate ballistic ions in comparable proportion to the existing population of Rydberg molecules.  

Selective field ionization experiments clearly show that these peripheral volumes evolve to a state in which electrons bind very weakly to a single ion in a high Rydberg orbital or in a quasi-free state in which they are held by the space charge of several ions.  

Despite evidence suggesting the presence of free electrons and readily available super-elastic collisional channels for electron heating, the $x$, $y$ and $z$ dimensions of separated volumes, measured as a function of time, establish that bifurcated plasmas expand very slowly.  Rates of expansion suggest electron temperatures near the correlation limit \cite{Niffenegger}:  Ion radial velocities reflect little more kinetic energy than might be released by the spatial correlation of ions.    

Bifurcated plasmas relax to exhibit remarkable chemical stability with respect to neutral fragmentation on a millisecond timescale.  Charge separation persists, despite ultrafast predissociation channels that exist for nitric oxide molecular Rydberg states with lower angular momentum at all principal quantum numbers, and efficient electron Rydberg inelastic collisions that rapidly scramble populations in $n$ and $\ell$ \cite{Saquet}.

\subsubsection{What is the state of the bifurcated ultracold plasma? } \label{state}

Our imaging experiment clearly establishes the existence of durable volumes that bifurcate from an elliptical Rydberg gas as it evolves to plasma.  Figure \ref{fig:traces} represents the expansion and decay observed for a typical case, in which the plasma forms with an estimated initial peak density of $3 \times 10^{11}~{\rm cm}^{-1}$.  Here, we find a state of arrested decay after an evolution time of 10 $\mu$s.  At this point, neglecting incipient bifurcation, images measured experimentally determine a Gaussian ellipsoid plasma volume with $\sigma_x$, $\sigma_y$, $\sigma_z$ principal axis dimensions of 1.0, 0.55 and 0.7 mm, and a peak density of $4 \times 10^{10}~{\rm cm}^{-1}$.  

SFI spectra, such as those displayed in Figure \ref{fig:SFI}, evolve for all initial principal quantum numbers to exhibit electron binding energy distributions that peak sharply in an appearance potential range from 0 to 10 V cm$^{-1}$.  This weakly bound electron signal persists as the dominant component in spectra measured after evolution times as long as 20 $\mu$s \cite{Haenel_width}.  Over this time interval, we find that the residual Rydberg signal, evident at low density in Figure \ref{fig:SFI}, relaxes to lower principal quantum number and all but disappears. 

The SFI spectrum tells us that the nitric oxide ultracold plasma predominantly binds electrons with an appearance potential of 10 V cm$^{-1}$ or less.  Images recorded in $x$, $y$ and $z$ show that this plasma expands very slowly, establishing that it redirects very little of its Coulomb potential energy to the radial motion of ions.  After an initial period of decay with a time constant of approximately 2 $\mu$s, the Rydberg molecules in quasi-equilibrium with NO$^+$ ions and electrons grow stable with respect to predissociation on a millisecond timescale.  It remains now to define an arrested state that best accounts for these properties of the system.  

The dominant SFI signal at low appearance potential reflects an abundance of high-Rydberg molecules ($n>80$) or quasi-free electrons of very low kinetic energy bound by the plasma space charge, or both.  A frozen high-$n$ Rydberg gas has an instantaneous dimensional stability and dissociative lifetime consistent with present observations, as does a fully ionized ultracold plasma with an electron temperature in the range of 5 K.  But, could either such state persist as long as a millisecond?

\begin{figure*}
\vspace{-5 pt}
\centering
\includegraphics[width= 1 \textwidth]{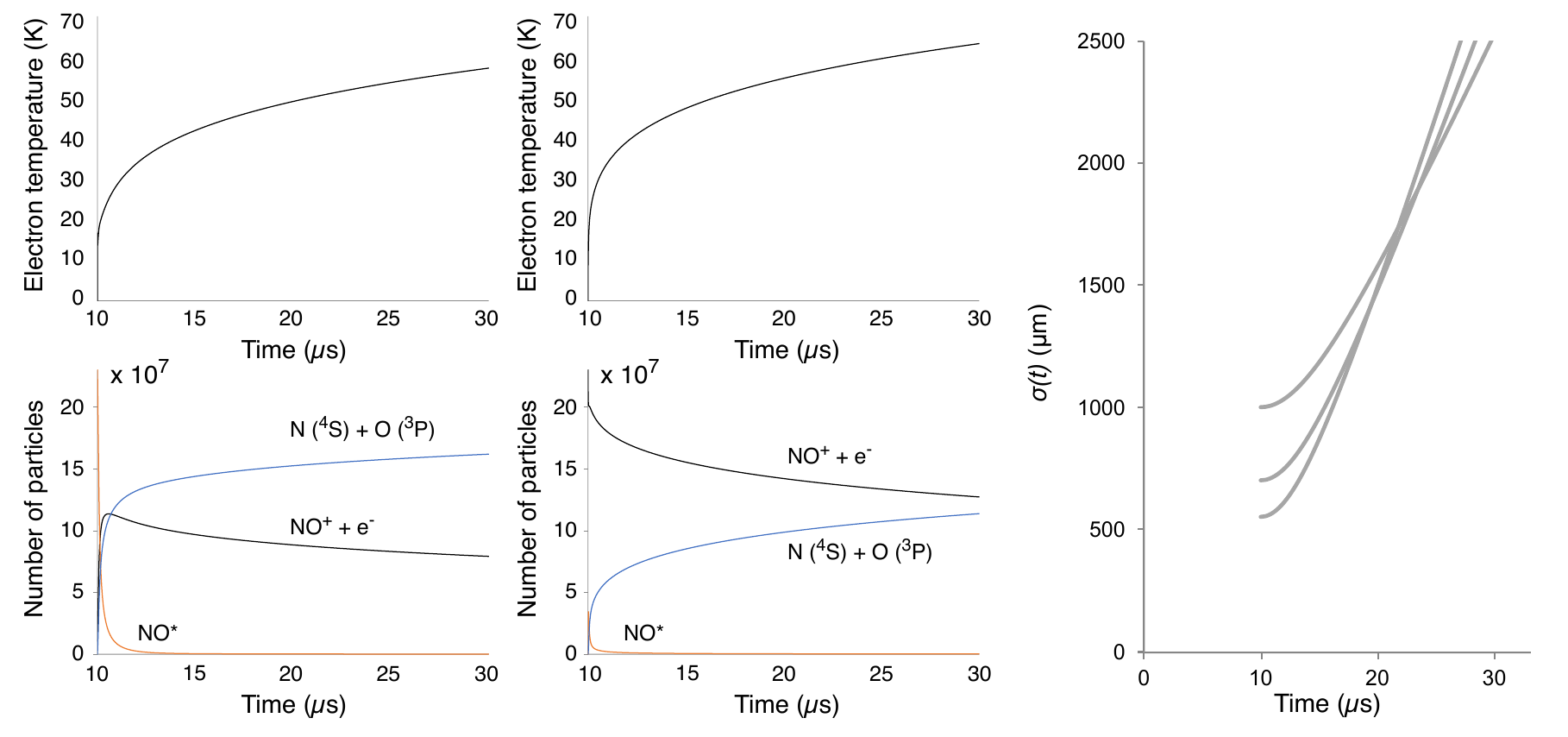}
   \caption{Results of coupled rate-equation simulations describing the relaxation of an $n_0=80$ Rydberg gas of NO (left) and an ultracold plasma of NO$^+$ ions and electrons with $T_e(0)=5$ K (centre).  Both simulations represent the initial density distribution of particles by a 5$\sigma$ Gaussian ellipsoid with principal axis dimensions, $\sigma_x=1.0$ mm, $\sigma_y=0.55$ mm, $\sigma_z = 0.7$ mm, as measured for the plasma represented in Figure \ref{fig:traces} after am expansion of 10 $\mu$s.  Evolution proceeds in 100 concentric shells enclosing set numbers of kinetically coupled particles, linked by a common electron temperature that evolves to conserve energy globally.  Upper frames describe the change in electron temperature as a function of time.  Lower frames show changes in the total number of ions and electrons, Rydberg molecules and neutral dissociation products N($^4$S) and O($^3$P).  The rightmost frame gives the results of a shell-model hydrodynamic simulation of the expansion of a Gaussian ellipsoid with the dimensions measured at 10 $\mu$s for the arrested plasma in Figure \ref{fig:traces}, assuming an electron temperature that rises to 60 K, with curves, reading from the bottom on the left, for  $\sigma_y(t)$, $\sigma_x(t)$ and $\sigma_x(t)$.   }
\label{fig:compilation}
\end{figure*}

Figure \ref{fig:rate_eqn} suggests on classical grounds that this is unlikely.  Here we see in coupled-rate equation simulations that a Rydberg gas prepared in $n_0=50$ with a density of $1 \times 10^{10}~{\rm cm}^{-1}$ evolves on a timescale of a few microseconds to form a plasma with an electron temperature that rises rapidly to exceed 100 K.  The ion-electron rate processes responsible for electron heating operate equivalently in a plasma model at this density configured to begin in a state of ions and free electrons with $T_e=5$ K.  Simulation models that account for dissociative recombination and predissociation predict plasma dissipation to neutral atoms on a $\sim 20$ $\mu$s timescale \cite{PCCP,Saquet}.  

However, the simulation conditions of Figure \ref{fig:rate_eqn}, as well as those explored in references \cite{PCCP} and \cite{Saquet} refer to particle distributions of uniform density.  Laser-crossed molecular beam illumination forms a Rydberg gas with a Gaussian ellipsoidal density distribution, measured after 10 $\mu$s of evolution to have representative dimensions as detailed above.  Measured along any Cartesian coordinate, the core of this distribution forms a sharp peak at a density of $4 \times 10^{10}~{\rm cm}^{-1}$.  But, particles with larger axial displacements occupy a much larger share of the three-dimensional ellipsoidal volume.  This extended region holds a substantial fraction of the total number of Rydberg molecules, ions and electrons under conditions of much lower density.  We must examine whether these rarified regions of the cold Rydberg gas significantly affect the lifetime predicted for an ultracold plasma of nitric oxide under conditions of classical avalanche and evolution.

For this purpose, we have developed a coupled rate-equation simulation model appropriate to the ellipsoidal density distribution measured for the arrested state of NO plasma that forms after an evolution time of 10 $\mu$s.  Our model uses 100 ellipsoidal shells to represent intervals of uniform density spanning a range of 5$\sigma$ in the ellipsoidal volume.  Each shell encloses a set number of particles, defined by its kinetically coupled sum of NO Rydberg molecules, NO$^+$ ions and electrons, and pairs of dissociation products N($^4$S) + O($^3$P).  The model assumes stationary molecules and ions.  We also assume that local quasi-neutrality confines electrons to their original shells, but their high thermal conductivity assures a uniform electron temperature.  

Prompt penning interactions ionize a defined fraction of Rydberg molecules in each shell and populations evolve according to Eqs. (\ref{level_i}) and (\ref{electron}).   Energy balance governs the change in electron temperature as defined by Eq. (\ref{energy}), evaluated with reference to electron and Rydberg densities determined globally.  

Let us now apply this semiclassical coupled rate-equation model to test the stability of the two possible arrest states suggested by the time-resolved SFI spectra.  Figure \ref{fig:compilation} shows coupled rate-equation simulation results that describe the evolution of temperature and numbers of particles for shell models with initial conditions chosen to represent Gaussian ellipsoidal volumes with the dimensions and peak densities measured experimentally after an evolution time of 10 $\mu$s.  Here we test whether the system at this point can be described as a Rydberg gas of high principal quantum number ($n=80$), or a fully ionized plasma of NO$^+$ ions and electrons with an electron temperature of 5 K.  

Figure \ref{fig:compilation} shows that neither starting point affords a dynamical model with the stability we see experimentally.  The classical model $n=80$ Rydberg gas avalanches to plasma in less than one microsecond.  Three-body recombination heats plasma electrons to 40 K on a timescale of a few microseconds, rising thereafter to more than 60 K.  Rydberg molecules deactivated in the initial avalanche dissociate in the first two microseconds.  On a longer timescale, direct dissociative recombination and three-body recombination followed by Rydberg predissociation feed a sustained, slower rate of decomposition to neutral atoms.  

In the opposite limit of an ultracold plasma modelled as a gas of ions and low electron temperature electrons, we find much the same classical evolution.  Initially, the large cross section for three-body recombination at low temperature consumes plasma ions and electrons, forming Rydberg molecules that promptly predissociate.  The electron gas heats, giving rise to a plasma steady state that evolves with the same dissipative kinetics as the model calculation beginning with a Rydberg gas.    

Most importantly, these rate-equation simulations predict that, in a classical regime, any ultracold plasma with the properties evident in the SFI spectra of Figure \ref{fig:SFI} must evolve to develop a high electron temperature, regardless of whether the dominant structure seen at low appearance potential signifies high-Rydberg molecules or electrons quasi-bound by the space charge, or both.  

The rising electron temperature called for by the coupled rate-equation calculation has significant hydrodynamic implications.  We can apply the shell model of Section \ref{evo} to predict the expansion rate of a plasma that undergoes classical relaxation.   Figure \ref{fig:compilation} on the right plots the growing principal axis widths of a Gaussian ellipsoid that begins with the dimensions of our plasma at 10 $\mu$s and an electron temperature predicted by the coupled rate-equation model.  This predicted rate of expansion far exceeds anything we have observed experimentally.  Table \ref{tab:expan} compares radial velocities in the ballistic limit as determined by the model calculation with the expansion rates of plasma volumes observed in the present experiment, independent of the selected principal quantum number or density of the initial Rydberg gas.  

\begin{table}[h!]
 \caption{Ballistic velocities in m s$^{-1}$ of 1) NO molecules in a volume of the molecular beam marked by its intersection with laser beams, $\omega_1$ and $\omega_2$, 2) Ions in the ultracold plasma that evolves from a Rydberg gas with $n_0=32$ formed at an initial density of $3 \times 10^{10}$ cm$^{-3}$, as determined by plasma image widths measured in the $x$, $y$ and $z$ dimensions as a function of time, 3) Limiting velocities determined by the shell-model hydrodynamic simulation results displayed in Figure \ref{fig:compilation}. } 
 \label{tab:expan}
 \vspace{5 pt}
 \begin{tabular}{lrrr} 
%  \multicolumn{4}{c}{\hspace{60 pt} Expansion velocity (m s$^{-1}$)} \\
 &$\partial_t \sigma_x$  &$\partial_t \sigma_y$ & $\partial_t \sigma_z$ \\
 \toprule 
% \end{tabular}
%\begin{tabular}{lrrr} 
  1) marked & 13 & 4 & 14 \\
  2) $n_0=32$ & 25 & 13 & 28 \\
  3) classical model & 103 & 150 & 130 \\
 \hline
\end{tabular}
%\vspace{-15pt}
\end{table}

In Figure \ref{fig:compilation}, model calculations describe a rate of plasma dissipation by neutral dissociation that slows over time.  We can understand this as a consequence of the increase in electron temperature that accompanies conventional avalanche.  A rising electron temperature suppresses both dissociative and three-body recombination, such that ion neutralization becomes the rate-determining step in plasma decay.  In a conventional plasma confined somehow to the very low electron temperature required by the experimental expansion rates we measure (Table \ref{tab:expan}), sustained dissociation to neutral products would occur at a high rate near that exhibited at very early evolution times in Figure \ref{fig:compilation}.

\subsubsection {Bifurcation as quench: Transition to a state of arrested relaxation}

The results of selective field ionization experiments, such as those pictured in Figure \ref{fig:SFI}, establish the state of the bifurcated plasma as one of high-Rydberg molecules and electrons weakly bound to a space charge of NO$^+$ ions.  We observe the prompt expansion of electrons and accelerated ions that initiates the process of bifurcation.  On simple electrodynamics grounds, these initial ions reach ballistic velocities in a few hundred nanoseconds, after travelling only a few tens of micrometers. This ambipolar transfer of the thermal energy of the electrons to the radial energy of the ions quenches $T_e$ to the order of $T_i$.  We can expect that $T_i$ changes slightly from the initial conditions of the ground-state NO in the molecular beam owing to ion-ion correlation heating and phase-space cooling associated with expansion.  

Ballistic ions with velocity components in $\pm x$ expand into the wings of the elliptical Rydberg gas.  Resonant charge exchange transfers ion momentum to Rydberg molecules.  This process recurs, creating a uniform $\pm x$ velocity field in which ions, cold electrons and Rydberg molecules all stream together.  For higher principal quantum numbers at higher Rydberg gas densities, the observed recoil velocity of bifurcating plasma volumes approaches half the initial radial velocity of the ambipolar accelerated core ions.  This suggests a medial expansion of ions and electrons that advances to meet an approximately equal density of cold Rydberg molecules.  

Internal energy of the system in all coordinates flows to the relative motion of bifurcating volumes,  quenching the internal state of the plasma.  As their relative velocities approach zero, ions and Rydberg molecules naturally move to positions of minimum potential energy.   These spatial correlations deplete the leading and trailing edges of the initially random distribution of nearest neighbours in the Rydberg gas \cite{Sadeghi:2014}, forming a random, three-dimensional network in which the distribution of NO$^+$/NO$^+$ distances, $r$, (referring both to Rydberg molecules and bare ions) peaks sharply at a Wigner-Seitz radius, determined by the density by $a_{ws}=(3/4\pi \rho)^{1/3}$.   The plasma thus self-assembles to form a correlated spatial distribution of intermolecular distances dictated entirely by internal forces.  

Bifurcation quenches this ensemble of ions and electrons to an annealed domain of low energy, perhaps to an ultracold regime.  Here, as evidenced by the SFI traces in Figure \ref{fig:SFI}, the ion-electron quasi-equilibrium populates a distribution of e$^-$-- NO$^+$ binding energies, just below and slightly above the isolated-molecule ionization threshold.  

Over time Rydberg-electron inelastic scattering processes ought to broaden this distribution \cite{PVS}, heating electrons and driving Rydberg molecules to lower $n$, where fast predissociation forms neutral fragment N and O atoms \cite{Remacle,Saquet}.  For the densities spanned by the SIF spectra in Figure \ref{fig:SFI}, classical simulations predict relaxation rates on a nanosecond to microsecond timescale \cite{Hung2014}.  However, as detailed in Section \ref{state}, properties of the bifurcated ultracold plasma, evident here for nearly a millisecond, show neither the expansion one would associate with electron heating nor dissipation of plasma charge by Rydberg relaxation and fragmentation to neutral atoms.  

High Rydberg systems have been widely studied in the context of light-matter interactions governed by exceedingly strong dipole-dipole interactions \cite{PilletJOSA,Hofmann}.  Laser excitation to Rydberg states gives rise to a number of important phenomena of current interest in quantum optics and many-body quantum dynamics, including entanglement, dipole blockade and electromagnetic-induced transparency \cite{Beterov,Saffman_JPB_rev,Sub_Poisson,Firstenberg}, as well as coherent energy transport, optical bistability and directed percolation \cite{Barredo,Carr,Marcuzzi,MarcuzziPRL}.

Long-time dynamics in the present system evolve in a similar landscape of dipole-dipole interactions with two important differences:  i) Electron-impact avalanche erases memory of the initial laser field.  As evident in Figure \ref{fig:SFI}, avalanche redistributes population over a complete manifold of high Rydberg states and quasi-bound electrons without reference to an electromagnetic field.  ii) Accordingly, no dipole blockade governs the minimum Rydberg-Rydberg distance.  In a dissipative system, the spatial coordinates of Rydberg molecules and ions relax to positions of minimum potential energy.  

Under these conditions, each Rydberg molecule undergoes an excitonic interaction with its nearest neighbour  \cite{SunRobicheaux,Deiglmayr2016,ParkGallagher,Buchleitner1}, either a Rydberg molecule in a different high-$n$, high-$l$ electronic state or an ion in the field formed by the quasi-continuum of electrons bound to multiple charge centres.  Importantly, every such interaction in the bifurcated plasma randomly pairs two molecules in \emph{different} excited states to define a unique, resonant close-coupled interaction.    

The bifurcated ultracold plasma shows an anomalous immobility in both the spatial coordinates of ions and electrons, as well as in the relaxation of electron binding energy to levels of rapid molecular predissociation.  However, the degree of charge mobility necessary to drive fast ambipolar expansion requires a substantial transfer of energy to electrons by the collisional relaxation of Rydberg molecules.  We could thus explain the very long lifetime of the plasma with respect to expansion and predissociation -- and the persistent structure of weak electron binding energy in the selective field ionization spectrum -- by an evolution suppressed solely in the coordinates of electron binding energy in the bifurcated plasma.  

Questions clearly remain:  Can a system strongly driven to quench and self-assemble form a glass or localize, at least in the coordinates of electron binding energy?  The range of conditions in this experiment yields systems with vastly different initial density and avalanche temperature.  Does relaxation to form arrested volumes with the same final density and internal energy signify a universal macroscopic physics?  Does a network of quantum mechanical interactions explain the slow long-time dynamics of these systems?  An incoherent mechanism may yet emerge, but our experimental observations substantially narrow the possibilities that can rely on conventional kinetics.

\begin{acknowledgements}

This work was supported by the US Air Force Office of Scientific Research (Grant No. FA9550-12-1-0239), together with the Natural Sciences and Engineering research Council of Canada (NSERC), the Canada Foundation for Innovation (CFI) and the British Columbia Knowledge Development Fund (BCKDF).  

\end{acknowledgements}

\bibliography{Image_refs,QMBL,AF_bib,Phys_Rev_A_(2017)}

\end{document}